\documentstyle[epsf,aps,preprint,tighten,floats]{revtex}
\begin{document}
\draft
\title{Gravitational waves from inspiraling compact binaries: \\
       Second post-Newtonian waveforms as search templates}
\author{Serge Droz and Eric Poisson}
\address{Department of Physics, University of Guelph, Guelph,
         Ontario, N1G 2W1, Canada}
\date{Submitted to Physical Review D, May 13, 1997}
\maketitle
\begin{abstract}
We ascertain the effectiveness of the second post-Newtonian 
approximation to the gravitational waves emitted during the
adiabatic inspiral of a compact binary system as templates for
signal searches with kilometer-scale interferometric detectors. 
The reference signal is obtained by solving the Teukolsky 
equation for a small mass moving on a circular orbit around 
a large nonrotating black hole. Fitting factors computed from 
this signal and these templates, for various types of binary
systems, are all above the 90\% mark. According to Apostolatos' 
criterion, second post-Newtonian waveforms should make acceptably 
effective search templates. 
\end{abstract}

\section{Introduction}

The current construction of kilometer-scale interferometric 
gravitational-wave detectors, such as the American LIGO (Laser 
Interferometer Gravitational-wave Observatory \cite{1}) and the 
French-Italian VIRGO \cite{2}, has motivated a lot of recent work 
on data analysis tools to search and measure the expected 
gravitational-wave signals \cite{3}. Thus far, the mostly 
studied waves have been those generated by the adiabatic 
inspiral of a compact binary system, composed of neutron stars 
and/or black holes \cite{4}. Because of their reliably large
event rate \cite{5,6,7}, these waves are currently believed 
to be the most promising for detection by LIGO/VIRGO. And because 
their waveforms can be accurately predicted, searching and measuring 
these waves will be best accomplished with the well-known technique 
of matched filtering \cite{8}. In this paper we focus on certain 
issues related to the {\it search} of gravitational waves from 
inspiraling compact binaries. 

The method of matched filtering \cite{8} relies on the fact that an 
accurate model can be constructed for the expected signal. The model 
(usually called the template) depends on various parameters, such as 
the signal's time of arrival, its initial phase, the masses of the 
companions, their spin angular momenta, and others. The 
template is cross-correlated with the detector output, and the 
parameters are varied until the largest correlation is obtained. 
If the maximum correlation (usually called the signal-to-noise 
ratio) is larger than a specified threshold value, then a signal 
is concluded to be present. 

A large effort is currently underway to calculate inspiraling binary
waveforms to high order in a post-Newtonian approximation \cite{9,10}. 
This approach is based on the assumption that the orbital motion is 
sufficiently slow that the waves can be expressed as a power series 
in $v/c$, the ratio of the orbital velocity to the speed of light. 
Because the series converges very slowly, if at all, a large number 
of terms are required \cite{11,12}. Thus far, the waveforms have been 
computed to order $(v/c)^5$ (2.5 post-Newtonian, or 2.5PN, order) 
beyond the leading-order, quadrupole-formula expressions \cite{10}.   

Central to the theory of matched filtering is Wiener's theorem \cite{8}, 
which states that the expectation value of the correlation is maximized 
when the template parameters match those of the actual signal. For the 
theorem to apply, however, the functional form of the template must be 
the same as that of the signal, so that template and signal are allowed 
to disagree only in the value of their parameters. In view of the fact 
that the signal is governed by the exact laws of general relativity, 
while the template  necessarily constitutes an approximation, we must 
expect that the actual signal-to-noise ratio will be somewhat less than 
the maximum allowed by Wiener's theorem \cite{12.5}. We must also expect 
that the value of the template parameters which maximize the correlation 
will not precisely match the value of the source parameters. 

Apostolatos \cite{13} has introduced the {\it fitting factor} (FF) as 
a measure of template imperfection. The fitting factor is the ratio of 
the actual signal-to-noise ratio, obtained with the imperfect template, 
to the signal-to-noise ratio that would be obtained if a perfect
template were available \cite{12.5}. A fitting factor of unity means 
that the template is a very accurate representation of the signal (a 
perfect template). A value less than unity means that the template 
reproduces the signal only imperfectly. The loss in event rate 
due to template imperfection is given by $1-\mbox{FF}^3$ \cite{13}. 

The fitting factor is computed as follows \cite{13}. Let $s(t)$ be 
the actual signal, and let $h(t;\bbox{\theta})$ be the templates, 
with the vector $\bbox{\theta}$ representing the parameters. We 
denote the Fourier transforms of these functions by $\tilde{s}(f)$ 
and $\tilde{h}(f;\bbox{\theta})$, respectively, where $\tilde{a}(f)
= \int a(t) e^{2\pi i f t}\, dt$ for any function $a(t)$.
We define the {\it ambiguity function} ${\cal A}(\bbox{\theta})$ by
\begin{equation}
{\cal A}(\bbox{\theta}) = \frac{(s|h)}{\sqrt{(s|s)(h|h)}},
\label{1}
\end{equation}
where, for any functions $a(t)$, $b(t)$,
\begin{equation}
(a|b) = 2 \int_0^\infty \frac{\bar{\tilde{a}}(f) \tilde{b}(f) +
\tilde{a}(f) \bar{\tilde{b}}(f)}{S_n(f)}\, df.
\label{2}
\end{equation}
Here, an overbar indicates complex conjugation, and $S_n(f)$
is the (one-sided) spectral density of the detector noise. For
interferometric detectors of the advanced-LIGO type, it is
appropriate \cite{14} to set $S_n(f) = S_0 
[ (f_{\rm min}/f)^4 + 2 + 2(f/f_{\rm min})^2 ]$ for 
$f > 10\ \mbox{Hz}$, while $S_n(f) = \infty$ for 
$f > 10\ \mbox{Hz}$; $S_0$ is a normalization constant
irrelevant for our purposes, and $f_{\rm min} = 70\ \mbox{Hz}$ 
is the frequency at which the noise is minimum. The quantity 
$(a|b)$ is usually interpreted as an inner product in the Hilbert 
space of gravitational-wave signals \cite{14}, and 
$\arccos {\cal A}(\bbox{\theta})$ is the angle between the 
signal $s(t)$ and the template $h(t;\bbox{\theta})$. The fitting
factor is given by
\begin{equation}
\mbox{FF} = \max_{\bbox{\theta}}\, {\cal A}(\bbox{\theta}),
\label{3}
\end{equation}
and $\arccos \mbox{FF}$ is the minimum value of this angle. 
Apostolatos \cite{13} has suggested that templates for which 
$\mbox{FF} > 90\%$ would be acceptably accurate, leading 
to a loss in event rate no larger than approximately 27\%. 

To compute the fitting factor one must first provide a reference
signal $s(t)$ and a parameterized set of templates $h(t;\bbox{\theta})$.
While post-Newtonian waveforms make an obvious choice of templates,
the selection of a reference signal is much more delicate, because
an exact representation of the gravitational waves is not available.
In all studies undertaken thus far \cite{13,15,16,17,18,19,20}, the 
reference signal was also chosen to be a post-Newtonian waveform, but 
one computed with more accuracy than the templates. (For example, the 
signal is taken to be a 2.5PN waveform, while the templates are chosen 
to be Newtonian, or 1PN, or 1.5PN, or 2PN waveforms.) This approach is 
flawed, because even a high-order post-Newtonian approximation may 
still be a poor representation of the true general-relativistic 
signal \cite{11}.

In this paper we make a different choice of reference signal which, we 
believe, bears a closer resemblance to the true general-relativistic
signal for binary systems with small mass ratios. This signal is
obtained by solving the Teukolsky equation \cite{21} for the 
gravitational perturbations produced by the circular motion of a small 
mass around a large nonrotating black hole \cite{22,23}. In this 
approach, no assumption is made regarding the size of $v$, the orbital 
velocity. This is one advantage over post-Newtonian theory. However, 
one disadvantage is the necessary restriction to binary systems with 
small mass ratios. 

\begin{table}[t]
\caption{The binary systems considered in this paper are listed
in the increasing order of their total mass. The individual masses
$m_1$ and $m_2$, and the chirp mass ${\cal M}_{\rm actual}$ are given 
in solar units. The mass-ratio parameter $\eta_{\rm actual}$ is 
dimensionless. All quantities are defined in the text.}
\begin{tabular}{ccccc}
System & $m_1$ & $m_2$ & ${\cal M}_{\rm actual}$ & 
$\eta_{\rm actual}$ \\
\hline \\
$1.4 + 1.4$ &  1.4 &  1.4 & 1.2188 & 0.2500 \\
$0.5 + 5  $ &  0.5 &  5.0 & 1.2322 & 0.0826 \\
$1.4 + 10 $ &  1.4 & 10.0 & 2.9943 & 0.1077 \\
$10  + 10 $ & 10.0 & 10.0 & 8.7055 & 0.2500 \\
$4   + 30 $ &  4.0 & 30.0 & 8.7340 & 0.1038 
\end{tabular}
\end{table}

Our fitting factors are therefore computed by taking $s(t)$ to be 
derived from the equations of black-hole perturbation theory, and 
by taking $h(t;\bbox{\theta})$ to be given by the so-called restricted 
second post-Newtonian approximation. (This terminology will be explained
below; we use 2PN waveforms because the formally more accurate 2.5PN 
waveforms are known to be a poorer representation of the true 
signal \cite{11}.) We consider the binary systems listed in Table I, 
which consist of three low mass-ratio systems ($0.5 M_\odot + 5 M_\odot$, 
$1.4 M_\odot + 10 M_\odot$, and $4 M_\odot + 30 M_\odot$, 
where $M_\odot$ is the solar mass) and two equal-mass systems 
($1.4 M_\odot + 1.4 M_\odot$ and $10 M_\odot + 10 M_\odot$). 
Although our analysis is not expected to be valid for the equal-mass 
systems, we nevertheless include them as an indication of the 
robustness of our conclusions. For all binary systems the companions 
are assumed to be nonrotating.

\begin{table}[t]
\caption{Searching perturbation-theory signals with second 
post-Newtonian templates. For each binary system, and for two 
types of signals, we list the values of ${\cal M}^*$, $\eta^*$, 
and $t^*$ corresponding to the global maximum of ${\cal A}(t^*,
{\cal M}^*, \eta^*)$. The fitting factor is also listed. All 
quantities are defined in the text. For all numbers the 
numerical uncertainty is $\pm 5$ in the last digit.}
\begin{tabular}{cccccc}
System & Signal & ${\cal M}^*_{\rm max}$ & $\eta^*_{\rm max}$ & 
$t^*_{\rm max}$ & FF \\
\hline \\
$1.4 + 1.4$ & one-mode  & 1.0033 & 0.401 & $ 0.3608 \times 10^{-4}$ 
& 94.7\% \\
$1.4 + 1.4$ & $l\leq 3$ & 1.0033 & 0.402 & $ 0.3562 \times 10^{-4}$ 
& 93.0\% \\
$0.5 + 5  $ & one-mode  & 1.0017 & 0.550 & $ 0.2345 \times 10^{-4}$ 
& 97.8\% \\
$0.5 + 5  $ & $l\leq 3$ & 1.0017 & 0.550 & $ 0.2421 \times 10^{-4}$ 
& 95.2\% \\
$1.4 + 10 $ & one-mode  & 1.0013 & 0.496 & $-0.2186 \times 10^{-3}$ 
& 95.5\% \\
$1.4 + 10 $ & $l\leq 3$ & 1.0013 & 0.496 & $-0.2154 \times 10^{-3}$ 
& 91.4\% \\
$10  + 10 $ & one-mode  & 1.0001 & 0.388 & $-0.1197 \times 10^{-3}$ 
& 98.4\% \\
$10  + 10 $ & $l\leq 3$ & 1.0001 & 0.388 & $-0.1179 \times 10^{-2}$ 
& 91.7\% \\
$4  + 30  $ & one-mode  & 0.9859 & 0.367 & $-0.5256 \times 10^{-2}$ 
& 95.4\% \\
$4  + 30  $ & $l\leq 3$ & 0.9859 & 0.367 & $-0.5192 \times 10^{-2}$ 
& 86.6\%
\end{tabular}
\end{table}

Our results, displayed in Table II, show that the 2PN waveforms
make acceptably accurate templates, in the sense of Apostolatos'
criterion \cite{13}: except for the large-mass system $4 M_\odot 
+ 30 M_\odot$, the fitting factors are all above the $90 \%$ mark. 
This is the main conclusion of this paper. We must point out that 
our analysis ignores such important issues as discrete template 
spacing \cite{24,25}, computing power, and spin-induced signal 
modulations \cite{13,26}. We leave such considerations for future 
work. Throughout the paper we work in units such that $G=c=1$.

\section{Reference signal}

The formalism of black-hole perturbation theory \cite{22} returns 
the following expression for the gravitational waves emitted by a 
mass $m_1$ in a fixed circular orbit around a nonrotating black 
hole of mass $m_2 \gg m_1$:
\begin{equation}
s_+(t) - i s_\times(t) \propto
\sum_{l=2}^\infty \sum_{m=-l}^l A_{lm}(v)\,
e^{-i m \Omega t}\, \mbox{}_{-2} Y_{lm}(\theta,\phi).
\label{4} 
\end{equation}
Here, $s_+(t)$ and $s_\times(t)$ are the two fundamental polarizations
of the gravitational waves, $A_{lm}(v)$ are mode amplitudes, and 
$\mbox{}_{-2} Y_{lm}(\theta,\phi)$ are spin-weighted spherical 
harmonics \cite{27}, with $\theta$ and $\phi$ the polar angles of the
gravitational-wave detector. The missing normalization constant is 
unimportant for our purposes. In Eq.~(\ref{4}), the orbital 
velocity $v$ and the orbital angular velocity $\Omega$ are related 
by $v^3 = M\Omega$, where $M = m_1 + m_2$ is the total mass. The mode 
amplitudes $A_{lm}(v)$ must be calculated separately for each value of 
$v$ by numerically integrating the Teukolsky equation \cite{23}.

Equation (\ref{4}) describes the waves produced by a mass moving on a
fixed orbit. This expression is easily generalized to describe an 
adiabatic inspiral: the orbital velocity $v$ now becomes a function 
of time, which is determined by
\begin{equation}
\frac{dv}{dt} = \frac{32 \mu}{5 M^2}\, v^{9}\, \frac{P(v)}{Q(v)}.
\label{5}
\end{equation}
Here, $\mu = m_1 m_2 / M$ is the reduced mass, $P(v) = 
(dE/dt)/(dE/dt)_N$ is the rate at which the gravitational waves 
remove orbital energy from the system, normalized to the 
quadrupole-formula expression $(dE/dt)_N = -(32/5)(\mu/M)^2 v^{10}$,
and $Q(v) = (dE/dv)/(dE/dv)_N = (1-6v^2)(1-3v^2)^{-3/2}$ gives the
differential relation between orbital energy and orbital velocity, 
normalized to the Newtonian expression $(dE/dv)_N = - \mu v$. The 
function $P(v)$ must be computed numerically \cite{23}. As a 
consequence of the time dependence of $\Omega = v^3/M$, the 
factor $\Omega t$ in Eq.~(\ref{4}) must be replaced by 
$\Phi(t) = \int^t \Omega(t')\, dt'$.

To simplify the expressions we assume that the detector is situated 
at $\phi = 0$, so that the spherical harmonics are functions of 
$\theta$ only. (Because the orbit is circular, this assumption
implies no loss of generality.) We also assume that the detector 
is oriented in such a way that it is sensitive only to the ``$+$'' 
polarization \cite{28}. Simple manipulations, using the relation 
$A_{l,-m} = (-1)^l \bar{A}_{lm}$ \cite{29}, then bring 
Eq.~(\ref{4}) to the form
\begin{equation}
s(t) \propto \sum_{l=2}^\infty \sum_{m=1}^l \mbox{Re}\Bigl[
A_{lm}(v) e^{-i m \Phi(v)}\Bigr]\, S_{lm}(\theta),
\label{6}
\end{equation}
where $s(t) \equiv s_+(t)$,
\begin{equation}
\Phi(v) = \frac{5M}{32\mu} \int^v \frac{Q(v')}{v'^6 P(v')}\, dv',
\label{7}
\end{equation}
and $S_{lm}(\theta) = \mbox{}_{-2} Y_{lm}(\theta,0) + 
(-1)^l \mbox{}_{-2} Y_{l,-m}(\theta,0)$. The relation between 
$t$ and $v$ is given implicitly by Eq.~(\ref{5}).

Equations (\ref{6}) and (\ref{7}) give the reference signal in
the time domain. Computation of the fitting factor, however, 
requires an expression in the frequency domain. We must therefore
Fourier transform Eq.~(\ref{6}), which we do in the stationary phase
approximation. Skipping all intermediary steps, we find
\begin{equation}
\tilde{s}(f) \propto \sum_{l=2}^\infty \sum_{m=1}^l 
\biggl[\frac{Q(v)}{m v^{11} P(v)} \biggr]^{1/2}\, A_{lm}(v)\,
e^{i m \Psi(v)}\, S_{lm}(\theta),
\label{8}
\end{equation}
where
\begin{equation}
\Psi(v) = \frac{5M}{32\mu} \int_{v_0}^v 
\frac{(v^3 - v'^3)Q(v')}{v'^9 P(v')}\, dv'.
\label{9}
\end{equation}
In these expressions, $f$ and $v$ are related by $v = (2\pi M f/m)^{1/3}$,
and $v_0$ is an arbitrary constant which sets the zero of the phase 
function $\Psi(v)$, which must also be evaluated numerically.  

In Eq.~(\ref{8}), the sum over $l$ must necessarily be truncated at
some finite value. The dominant part of the signal comes from the  
$l=m=2$ mode. We shall refer to a signal containing only this mode 
as a {\it one-mode signal}. The second largest contribution to the 
signal comes from the $l=m=3$ mode, and in fact, all additional 
contributions can be neglected. We shall therefore truncate the sum 
at $l=3$, having numerically verified that additional terms are indeed 
irrelevant. Because $S_{33}/S_{22} = - (21/10)^{1/2}\sin\theta$, the
relative importance of the $l=3$ terms is maximized when $\theta = 
\pi/2$. We shall therefore set $\theta$ to this value throughout the 
paper. 

The signal must be cut off when the orbiting mass reaches the
innermost stable circular orbit at $v=6^{-1/2}$. Because the
$l=m=2$ mode dominates the signal, we simply let $\tilde{s}(f)=0$
when $f > (6^{3/2}\pi M)^{-1}$.

\section{Templates}

The restricted second post-Newtonian approximation to the
gravitational-wave signal is given in the frequency domain 
by \cite{30}
\begin{equation}
\tilde{h}(f;\bbox{\theta}) \propto f^{-7/6}\, 
e^{i\psi(f)},
\label{10}
\end{equation}
where $\psi(f) = 2\pi f t_c - \phi_c + \phi(f)$,
\begin{eqnarray}
\phi(f) &=& \frac{3}{128} 
(\pi {\cal M} f)^{-5/3} \Biggl[1 + \frac{20}{9}  
\biggl( \frac{743}{336} + 
\frac{11}{4} \eta \biggr)\, x^2 - 16\pi x^3 
\nonumber \\ & & \mbox{}
+ 10 \biggl( \frac{3058673}{1016064} + \frac{5429}{1008}\eta 
+ \frac{617}{144} \eta^2 \biggr)\, x^4 \Biggr],
\label{11}
\end{eqnarray}
with $x^3 \equiv \pi \eta^{-3/5} {\cal M} f$. Here, 
$\{\phi_c, t_c, {\cal M}, \eta\}$ are the
template parameters, with $\phi_c$ the phase at
coalescence (formally $f=\infty$), $t_c$ the time at coalescence, 
$\cal M$ the chirp mass, and $\eta$ the mass-ratio parameter. 
The term ``restricted'' refers to the fact that while the phase 
function $\phi(f)$ is calculated to 2PN order beyond the 
quadrupole-formula expression, the amplitude function 
$f^{-7/6}$ is kept at leading order.

Because the templates do not have the same functional form as 
the signal, $\cal M$ and $\eta$ do not have direct physical meaning. 
However, {\it if} the templates were an accurate description of the 
waves, then the signal would be best reproduced when ${\cal M} = 
{\cal M}_{\rm actual}$ and $\eta = \eta_{\rm actual}$, where \cite{30}
\begin{equation}
{\cal M}_{\rm actual} = \frac{(m_1 m_2)^{3/5}}{(m_1 + m_2)^{1/5}},
\qquad
\eta_{\rm actual} = \frac{m_1 m_2}{(m_1 + m_2)^2},
\label{12}
\end{equation}
with $m_1$ and $m_2$ denoting the actual masses of the companions.
It is therefore convenient to introduce the rescaled parameters
${\cal M}^*$ and $\eta^*$, defined by
\begin{equation}
{\cal M} = {\cal M}^* {\cal M}_{\rm actual}, \qquad
\eta = \eta^* \eta_{\rm actual}.
\label{13}
\end{equation}
It is also convenient to make the following transformations:
\begin{equation}
\phi_c = \phi_0 + \phi^*, \qquad
t_c = t_0 ( 1 + t^* ),
\label{14}
\end{equation}
where
\begin{equation}
\phi_0 = 2\pi f_0 t_c + \phi(f_0), \qquad
t_0 = - \frac{1}{2\pi}\, \frac{d\phi}{df}(f_0),
\label{15}
\end{equation}
with $f_0$ an arbitrarily chosen frequency. Their purpose is to 
ensure that when $\phi^* = t^* = 0$, then $\psi(f_0) = 
d\psi/df(f_0) = 0$. These relations are analogous to those 
holding for $\Psi(v)$ at $v=v_0$ [see Eq.~(\ref{9})]. We shall 
henceforth take $\bbox{\theta} = (\phi^*,t^*,{\cal M}^*,\eta^*)$ 
to be our template parameters.

\section{Computing the fitting factor}

Evaluation of the ambiguity function, Eq.~(\ref{1}), for a given 
binary system and a given set of template parameters, 
is a straightforward numerical problem. First $A_{lm}(v)$ and 
$P(v)$ are computed once and for all, and tabulated for many 
values of the orbital velocity. Next, $\Psi(v)$ is calculated 
for the binary system under consideration. Combining the results 
and summing over the relevant modes gives $\tilde{s}(f)$, 
Eq.~(\ref{8}). On the other hand, $\tilde{h}(f;\bbox{\theta})$ 
is given in analytical form by Eqs.~(\ref{10})--(\ref{15}). 
The overlap integral $(s|h)$ is then evaluated. To do this 
efficiently it is advantageous to choose the constants $f_0$ and 
$v_0$ [see Eqs.~(\ref{9}) and (\ref{15})] so that $\Psi(f)$, 
$\psi(f)$, and their slopes all vanish where the integrand is largest. 
This ensures that signal and template are approximately in phase in 
the frequency interval that contributes the most to the integral. Now, 
the integrand is fairly well approximated by $|\tilde{h}|^2/S_n(f) 
\propto z^{-7/3}/(z^{-4} + 2 + 2z^2)$, where $z \equiv f/f_{\rm min}$, 
which is maximum at $z \simeq 0.6654$.
We therefore set $f_0 = 46.58\ \mbox{Hz}$. And since $\tilde{s}(f)$ 
is dominated by the $l=m=2$ mode, we correspondingly put 
$v_0 = (\pi M f_0)^{1/3}$. With these choices, computation of $(s|h)$
is optimized. Evaluation of $(s|s)$ and $(h|h)$ presents no additional 
difficulties, and combining the results gives ${\cal A}(\bbox{\theta})$.

\begin{figure}[t]
%
  \protect\centering
  \mbox{ \epsffile{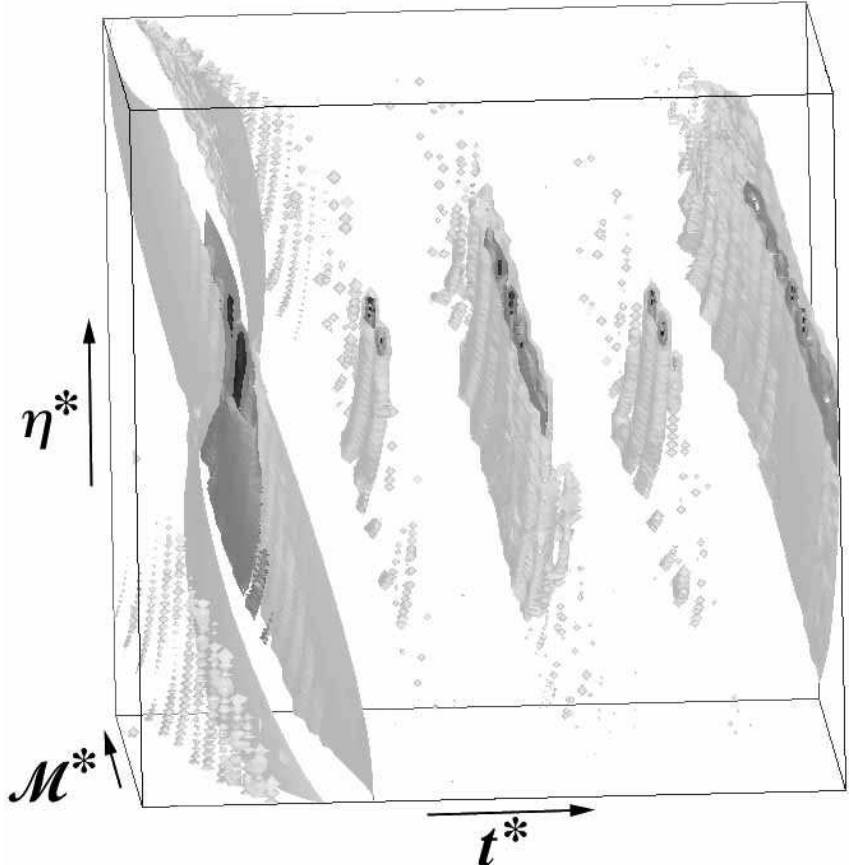}}
%
\caption{Shown here are surfaces of constant ambiguity function 
${\cal A}(t^*,{\cal M}^*,\eta^*)$ for the binary system $1.4M_\odot 
+ 10 M_\odot$. The reference signal is a one-mode signal, and the
templates are second post-Newtonian waveforms. The light grey surfaces 
represent ${\cal A} = 15\%$, the dark grey surfaces represent ${\cal A} 
= 30\%$, and the black surfaces represent ${\cal A} = 50\%$. The global 
maximum of 95.5\% is contained within the large structure to the left of 
the box. The other structures contain additional local maxima, and these 
make an automated search for the true maximum difficult. The region
of parameter space covered by the box is $-0.06 < t^* < 0.35$,
$0.96 < {\cal M}^* < 1.05$, and $0.35 < \eta^* < 0.90$.}
\end{figure}

We now discuss the issue of maximizing the ambiguity function
over the template parameters. Maximization over $\phi^*$ is 
straightforward. Since $\phi^*$ comes as an overall phase in 
$\tilde{h}(f;\bbox{\theta})$, we have that $(s|h) = 
\mbox{Re}(|I| e^{i\alpha})$, where $\alpha$ is linear in $\phi^*$, 
and where $I$ is some function of the remaining parameters. 
Maximization over $\phi^*$ is achieved by making $\alpha = 0$, which 
yields $(s|h) = |I|$. Since $(h|h)$ and $(s|s)$ do not depend on 
$\phi^*$, we have also maximized ${\cal A}(\bbox{\theta})$ over 
$\phi^*$. We will denote this reduced ambiguity function by 
${\cal A}(t^*,{\cal M}^*,\eta^*)$. Maximizing this over the remaining
parameters is {\it not} a straightforward numerical problem, because 
the ambiguity function displays a very rich structure, featuring many 
local minima and maxima (see Fig.~1). We proceed as follows. For each 
binary system under consideration we evaluate the reduced ambiguity 
function in a fairly large neighborhood of $(t^*,{\cal M}^*,\eta^*) = 
(0,1,1)$. The function is then displayed graphically with the help of 
a three-dimensional visualization software \cite{31}, 
and the approximate position of the global maximum is determined. 
Finally, the maximum is determined accurately by applying Powell's 
method as described in the book {\it Numerical Recipes} \cite{32}. 
The value of ${\cal A}(t^*,{\cal M}^*,\eta^*)$ at the global maximum 
is the fitting factor FF.

The calculations described in this section require a great deal of
numerical work. To minimize the possibility of making coding errors
which would affect our results and conclusions, a separate code
was written by each of the authors. Because the outputs of the two 
separate codes agree within the numerical uncertainty quoted in the 
caption of Table II, we are quite confident in the accuracy of our 
results. 

\section{Results and discussion}

Our results are displayed in Table II. For each binary system we
consider two types of signals. The one-mode signal contains only
the $l=m=2$ mode, while the $l\leq 3$ signal contains all modes
up to $l=3$. Inclusion of additional modes does not change the
results within the numerical uncertainty. We see that the fitting 
factors for the one-mode signals are large, all above approximately 
95\%. However, inclusion of additional modes decrease the 
fitting factors appreciably: the reduction ranges from approximately 
3\% for low-mass systems to approximately 7\% for large-mass systems.
We have verified that most of this reduction can be attributed to the
$l=m=3$ mode. Nevertheless, the fitting factors are all larger 
than 90\%, except for the large-mass system $4 M_\odot + 30 M_\odot$. 
The second post-Newtonian templates therefore meet the Apostolatos 
criterion \cite{13}, and we conclude that they should be adequate 
to search for gravitational waves emitted during the adiabatic 
inspiral of a compact binary system. 

\begin{table}[t]
\caption{Searching a one-mode signal with Newtonian templates.
For each binary system we list the values of ${\cal M}^*$ and 
$t^*$ corresponding to the global maximum of ${\cal A}(t^*,{\cal M}^*)$. 
The fitting factor is also listed. The numerical uncertainties are as 
in Table II.}
\begin{tabular}{cccc}
System & ${\cal M}^*_{\rm max}$ & $t^*_{\rm max}$ & FF \\
\hline \\
$1.4 + 1.4$ & 0.9944 & $-0.4120 \times 10^{-3}$ & 79.2\% \\
$0.5 + 5  $ & 0.9999 & $-0.3597 \times 10^{-4}$ & 51.6\% \\
$1.4 + 10 $ & 1.0201 & $-0.1057 \times 10^{-2}$ & 55.8\% \\
$10  + 10 $ & 1.0563 & $-0.1361 \times 10^{-1}$ & 70.1\% \\
$4   + 30 $ & 1.1503 & $-0.2339 \times 10^{-1}$ & 61.3\%
\end{tabular}
\end{table}

For comparison, we show in Table III the results obtained when
Newtonian templates are used. [We discard all post-Newtonian
corrections in Eq.~(\ref{11}), and $\eta^*$ disappears from the 
list of template parameters.] Here the fitting factors are computed 
using a one-mode signal, as this gives the largest result. We see that
these fitting factors are all much smaller than those of Table II. The
Newtonian templates do not meet the Apostolatos criterion \cite{13}.

The reduction in fitting factor that occurs when a one-mode signal 
is replaced by a $l\leq 3$ signal is clearly caused by the template's 
inability to keep track of the additional frequency components. We
have attempted to improve the performance of our templates by adding
a term
\begin{equation}
\tilde{h}_3(f;\bbox{\theta}) \propto \frac{3^{7/3}}{2^{10/3}}\, 
(1-4\eta)^{1/2}\, f^{-7/6}\, x\, \sin\theta\,
e^{i[2\pi f t_c - 3\phi_c/2 + \phi(2f/3)]},
\label{16}
\end{equation}
where the constant of proportionality is the same as in Eq.~(\ref{10}).
While $\tilde{h}(f;\bbox{\theta})$ describes waves at twice the orbital
frequency and is the leading-order term in a post-Newtonian expansion, 
$\tilde{h}_3(f;\bbox{\theta})$ describes waves at three times the orbital 
frequency and is a correction term of order $v$. Equation (\ref{16}) was 
obtained by Fourier transforming Eq.~(3b) of Ref.~\cite{9}. Our 
expectation was that this additional term would do a good job at 
reproducing the behavior of the $l=m=3$ mode, and that the loss in
FF would be regained. Calculating fitting factors with $h + h_3$ as 
templates shows otherwise. (Some thought must be given to the fact 
that $\phi^*$ no longer appears as an overall phase in the templates.) 
The improvement in FF was less than 1\% for the large-mass systems, 
and less than 0.1\% for the low-mass systems. This prompts us to conclude 
that it would be pointless to introduce additional frequency components 
in the search templates. 

It is interesting to observe that for low-mass systems, the maximum 
of the ambiguity function is obtained when ${\cal M}$ is approximately 
0.1\% above its actual value, while $\eta$ is approximately 40\% of its 
actual value. These numbers can be loosely interpreted as the systematic 
errors incurred when attempting to estimate the source parameters with 
imperfect 2PN templates. The systematic errors should be compared with 
the anticipated statistical uncertainties associated with the measurement 
of a noisy signal \cite{14,20,30,33}. Cutler and Flanagan \cite{14} have 
determined that for low-mass systems known to be composed of nonrotating 
companions, the statistical uncertainty associated with the estimation of 
${\cal M}$ is of the order of 0.005\%, while it is of the order of 1\% 
for $\eta$ \cite{34}. (This assumes detection with a signal-to-noise 
ratio of 10.) Second post-Newtonian templates therefore give rise to 
systematic errors that are much larger than the statistical 
uncertainties. While these templates could be adequate to search for 
signals, they should not be used as measurement templates.

\section*{Acknowledgments}

This work was supported by the Natural Sciences and Engineering
Research Council of Canada.


\begin{references}
\bibitem{12.5}
   Here and below, the signal-to-noise ratio is defined to be
   the expectation value of the maximum correlation between the
   noisy detector output and the templates.
\bibitem{34}
   Cutler and Flanagan derived these results using a 1.5PN waveform.
   We have recalculated the statistical uncertainties using a 2PN 
   waveform, and have found the same answer.
\end{references}
\end{document}